\documentclass[conference,11pt]{IEEEtran}
\makeatletter
\def\ps@headings{%
\def\@oddhead{\mbox{}\scriptsize\rightmark \hfil \thepage}%
\def\@evenhead{\scriptsize\thepage \hfil \leftmark\mbox{}}%
\def\@oddfoot{}%
\def\@evenfoot{}}
\makeatother
\usepackage{comment}
\usepackage{multirow}
\usepackage{amsmath,amssymb,amsfonts}
\usepackage{algorithmic}
\usepackage{graphicx}
\usepackage{textcomp}
\usepackage{xcolor}
\def\BibTeX{{\rm B\kern-.05em{\sc i\kern-.025em b}\kern-.08em
    T\kern-.1667em\lower.7ex\hbox{E}\kern-.125emX}}

\begin{document}

\title{Overview of Security of Virtual Mobile Networks}

\author{\IEEEauthorblockN{Ijaz Ahmad, Ilkka Harjula, Jarno Pinola}
\IEEEauthorblockA{\textit{VTT Technical Research Center of Finland, Espoo, Finland)}\\
Espoo, Finland \\
firstname.lastname@vtt.fi}
}
\maketitle

\renewcommand{\baselinestretch}{1.05}

\begin{abstract}

5G is enabling different services over the same physical infrastructure through the concepts and technologies of virtualization, softwarization, slicing and cloud computing. Virtual Mobile Networks (VMNs), using these concepts, provide an opportunity to share the same physical infrastructure among multiple operators. Each VMN Operator (VMNO) can have own distinct operating and support systems. However, the technologies used to enable VMNs have their own explicit security challenges and solutions. The integrated environment built upon virtualization, softwarization, and cloudification, thus, will have complex security requirements and implications. In this vain, this article provides an overview of the security challenges and potential solutions for VMNs.



\end{abstract}

\begin{IEEEkeywords}
Security; VMNs security; 5G security
\end{IEEEkeywords}



\section{Introduction}

With the advent of new technological developments in 5G such as virtualization and network slicing, Virtual Mobile Networks (VMNs) will provides an opportunity to share the same physical infrastructure among multiple operators. Each VMN Operator (VMNO) can have its own operating and support systems, service offering and user base. Being virtual in nature, the network can be easily scaled up and down as the need arises~\cite{8703474}. Due to vast array of benefits, since 2012 the European Telecommunications Standards Institute (ETSI) hosted the industry specification group for Network Function Virtualization (NFV) to apply the mainstream virtualization techniques to standardized network elements. Hence, parts or whole of the access, backhaul and core networks can be virtualized. 

NFV enables telecom operators to use commercial-of-the-shelf (COTS) general purpose network equipment to satisfy various needs with less costs compared to specialized purpose-specific dedicated hardware. Hence, NFV paved the path to separating network functions from purpose-built devices to be implemented in software which could be deployed on general purpose COTS equipment~\cite{YI2018212}. This capability of NFV has brought the flexibility and agility of clouds to communication networks in terms of facilitating different services on network equipment~\cite{CERRATO2015380}. Furthermore, NFV facilitates dynamic service creation and management in different network perimeters with much higher flexibility than legacy techniques~\cite{7045396}.

Software Defined Networking (SDN) facilitates virtualized network functions (VNFs) to be placed in different network perimeters~\cite{7194059}, and thus, NFV and SDN have become highly complementary. A VNF can be any function performed by a network node implemented solely in software. Hence, networking and service functions implemented as VNFs can be upgraded, chained, deployed, re-deployed or removed instantaneously. However, each of these technological enablers of VMNs, such as SDN, NFV, and clouds have their own security challenges and solutions~\cite{8334918}. Even though a lot of work has been done on security of each distinct technology, very little attention has been paid to the security of the integrated VMN environment. Therefore in this article, the security challenges, potential solutions, and existing security gaps of VMNs are discussed. This article is organized as follows: Section II discusses the security challenges and possible solutions for security of VMNOs along with the enabling technologies used in this direction. Section III discusses security management in VMNs along with trust, privacy and standardization efforts. Section IV briefly highlights the potential existing challenges and the article is concluded in Section V.

\section{VMN Security Challenges and Solutions}

VMNs will use 5G as the underlying infrastructure, and leverage the concepts of cloud computing besides NFV and SDN to efficiently place network functions, scale up resource for different functions when needed, and provide unified platforms for network management resulting in Telecommunication network as a Service (TaaS). Therefore, the security of VMNs will be dependent on security of 5G as a whole, and SDN, cloud platforms, and most importantly, virtualization technologies and NFV in parts. Therefore, we discuss first the security of 5G in brief and then security of each enabling technology below.

\subsection{Brief Overview of 5G Security Challenges}

The Next Generation Mobile Networks (NGMN) consortium suggests 5G to provide more than hop-by-hop or radio bearer security, which were common in 4G and prior generations of cellular networks~\cite{8712553}. Due to the inclusion of diverse services and technologies the security threat landscape will be much more different and complex. The most important challenges, that are more threatening in the case of 5G compared to previous generations~\cite{8712553, 8334918}, are summarized in Table~\ref{table1}.

\begin{table} [ht]
\scriptsize
\caption{Main security challenges faced by 5G}
\label{table1} 
\centering 
\begin{tabular}{|p{2 cm}  |p{6 cm} | } 
\hline 
\bfseries Security Challenge & \bfseries Description   \\ \hline

Denial of Service (DoS) Attack & Targeting availability of resources with DoS attacks on the infrastructure and end-user devices. \\ \hline
Flash network traffic & Sudden arrival of large number of packets causing a jamming like service breaks.  \\ \hline
Security of interfaces& Security lapses regarding interface encryption keys generated in home network and sent to visited network over insecure links.  \\ \hline
User plane integrity & Lack of cryptographic integrity protection for the user data plane in cases where the traffic terminates beyond mobile networks. \\ \hline
Lack of assurance &  Security lapses occurring due to lack of security assurance in multi-operator environments.  \\ \hline
Roaming security & Conflicts among multiple operators regarding subscriber-level security policies during roaming, usually requiring operators to share information.  \\ \hline

\end{tabular}
\vspace{-4mm}
\end{table}

The security of 5G is dependent on that of SDN, NFV, cloud platforms, and have characteristics: i) Supreme built-in-security following the principle of security-by-design for emerging systems and services, ii) Flexible security mechanisms leveraging the principles of NFV and SDN for deploying dynamic security functions, and iii) Automation leveraging AI for minimal human intervention. As a baseline,the 3GPP has defined a security architecture with entities outlined in Table II, that could also solve the challenges mentioned in Table I.

\begin{table} [ht]
\scriptsize
\caption{Main points of 3GPP security architecture}
\label{table2} 
\centering 
\begin{tabular}{|p{2 cm}  |p{6 cm} | } 
\hline 
\bfseries Security type & \bfseries Description (solution for challenges in Table I)   \\ \hline

Network access security & For secure authentication and access to network services. (User plane integrity ) \\ \hline
Network domain security & For secure exchange of signaling and user plane data among network nodes. (Security of interfaces) \\ \hline
User domain security & For secure user access to a user equipment (UE). (User plane integrity) \\ \hline
Application domain security & To enable user and provider domain applications to securely exchange messages. (User plane integrity, security assurance, roaming security) \\ \hline
Service-based architecture domain security & For network element registration, discovery, authorization, and service-based interfaces. \\ \hline
Visibility and configurability & Security features that inform users of security features operations. (DoS, security assurance)  \\ \hline

\end{tabular}
\vspace{-4mm}
\end{table}

\subsection{Security of SDN}

SDN is one of the main enabling technologies of VMNs since it provides abstractions of the physical network infrastructure~\cite{7901481}. SDN introduces network programmability and centralizes the network control to SDN controllers, thus, the security concerns mostly relate to these features~\cite{7226783}. For example, SDN enables applications to program or change the behavior of the network. This gives rise to the need of strong authentication and authorization techniques for applications. Moreover, the centralized controllers are favorable targets for DoS and resource exhaustion attacks. Fingerprinting the controllers, for example through time stamps of live packets in the network~\cite{azzouni2016fingerprinting}, or round-trip time~\cite{7480416} have been demonstrated. Therefore, devolving controller functions (e.g., local decision-making), hierarchical controllers, resilience through increased capabilities, and intelligent security systems using machine learning for proactive measures have been proposed~\cite{7226783}. 

SDN can be also used to improve the security of virtual networks~\cite{7399014}. Virtual Machine (VM) migration techniques using SDN can help to move resources to secure perimeters. For instance, live VM migration if the network is under a DoS attack can efficiently help in scalability through monitoring the load states (e.g. packet counter values) in the SDN forwarding plane. Live VM migration in legacy networks has been difficult for two reasons. One, network state unpredictability, and second, VM migration is limited to LAN since IP does not support VM mobility without session breakups. SDN solves these challenges through centralized programmable network control having visibility of global network state and independence of the layered IP stacks. Therefore, SDN can improve the security of VMNs.


\subsection{Security of Cloud Platforms}

Cloud computing~\cite{hayes2008cloud} has become a central part of mobile networks for a number of benefits ranging from radio access networks (RAN) to core networks~\cite{6963805}. Cloud computing concepts have been extended to meet latency requirements through edge computing~\cite{7807196, 8016573}, MEC~\cite{7931566}, and fog computing~\cite{7513863}. Virtualization of the cloud platforms for enabling novel services have many benefits of costs and efficiency. However, there are inherent security challenges in cloud platforms that are highly important when it comes to virtual systems on cloud platforms. For example, MEC suffer from latency during authentication, and the existing authorization, accounting and access control are not suitable for MEC leaving space to threats, as discussed in~\cite{8951281}. Therefore, novel techniques for security~\cite{8951281} and privacy~\cite{9004559} in MEC platforms must be adopted, and novel lightweight techniques need to be designed for fog platforms.


The two main inter-junction points of cloud and virtualization for wireless networks are cloud RAN (C-RAN)~\cite{6897914} and cloud-based core networks~\cite{7045396}. A survey on C-RAN security~\cite{7954591} outlines the main challenges and potential solutions. The existing challenges include the lack of universal C-RAN security framework, secure sensing techniques, trust and privacy, and the infancy of physical layer security. Furthermore, C-RAN pools Baseband units from multiple base stations into a centralized pool for statistical multiplexing gain~\cite{6897914}. Such centralization would invite DoS and other resource exhaustion attacks. On the core network side, most of the security challenges are related to signaling storms, DoS attacks, and the security dependability on SDN and NFV~\cite{ahmad20195}.

\subsection{Security of Virtual Machines}

A VM might be running one or several different VNFs. Thus, the security of virtual systems, specific to the VMNs, is multi-pronged. The security of virtual systems in VMNs revolve around hypervisor, VMs, and VNFs. There is also the concept of virtualized threats that refers to attacks against availability, integrity and confidentiality of software and hardware in VMN. All the VMs and hypervisors must be adequately secured from unauthorized access, change, and other disturbances. In VMNs, the hypervisor is a central entity that is not directly connected to users, thus the security threats arise from VMs. Therefore, similar to other centralized or core elements, the hypervisor must be protected through proper authentication, authorization and accountability mechanisms. Similarly, security mechanisms needed for availability must be in place since the unavailability of the hypervisor would be a serious problem for all services. The reliability of hypervisor requires security of VMs. Strong isolation mechanisms will be required to minimize the effects of malicious VMs on one another and on the hypervisor~\cite{liyanage2018comprehensive}. Defining and setting different security zones and traffic separation can also improve isolation-based security of VMs. However, most of these security approaches are yet to be seen due to the limited deployment of VMNs.

\subsection{Security of VNFs}

The concept of NFV to implement networking functions in software to be deployed on commodity network equipment led to the rise of VNFs~\cite{YI2018212}. Soon novel verticals will span multiple operator environments in the form of VNFs. Thus, VNFs can have a diverse threat vector. The security threats can arise from the software implementations, VNF configurations, security weaknesses in hypervisors and cloud platforms, as well as direct attacks on VNFs such as side-channel attacks, flooding attacks, and malware injection~\cite{8334918}. Due to the dynamic nature of VNFs, trust management is another serious concern since VNFs will be capable to move between multiple networks, and cloud platforms maintained by different owners and operators~\cite{8334918}. The targets of such attacks include user traffic, VNF code and policy input, and state of VNFs. Such attacks can be materialized by exploiting inherent limitations in operating environments including its software and hardware~\cite{8806681}. 

Similarly, serious security challenges can arise from interfaces, mainly when standardized interfaces are not defined~\cite{7502434}. Furthermore, the VNF package security validation check is highly important to avoid introducing security vulnerabilities in the whole system. Therefore, there are several proposals for confidentiality check through proper authentication and integrity verification for VNF packages onboarding into NFV systems. There are also other proposals for ensuring security of systems from malicious VNFs. For example, authors in~\cite{NFVOrchSec} propose and demonstrate a verification system for security attributes of different VNFs to protect NFV infrastructure (NFVI) using standard TOSCA~\cite{tosca2015tosca} data models.

\section{Security Management in VMNs}

Due to the dynamic nature of NFVI and VNFs, security management is highly complex VMNs. The complexity is due to consistent maintenance and management of VNF configurations and seamless transfer of state information from one VNF to another~\cite{7014200}. Similarly, the elasticity of NFV brings forth challenges in decomposing services for data and control planes, enforcing policies, and managing and controlling the entire network where control signals must go only through the trusted functional blocks such as VNF managers, VIM, and NFV orchestrator~\cite{7113220}. 

The ETSI specification release 3, security management and monitoring specification~\cite{ETSI2}, provides important insights into security management and monitoring problems. It states that traditional security systems will not scale for NFV, may result in inconsistent policies, inefficient processes and increase overall complexity. Monitoring in NFV deployments is highly complicated due to possibly concealed interfaces by consolidated verticals, functional silos, and collapsed stacks like shared memory and virtual sockets. In large-scale deployments, probing for security monitoring is complicated by the myriad of VNFs, vendor-proprietary implementations, and non-3GPP standardized interfaces, as well as automation and live migrations. 

\begin{figure}[ht]
  \centering
  \includegraphics[width=0.30\textwidth]{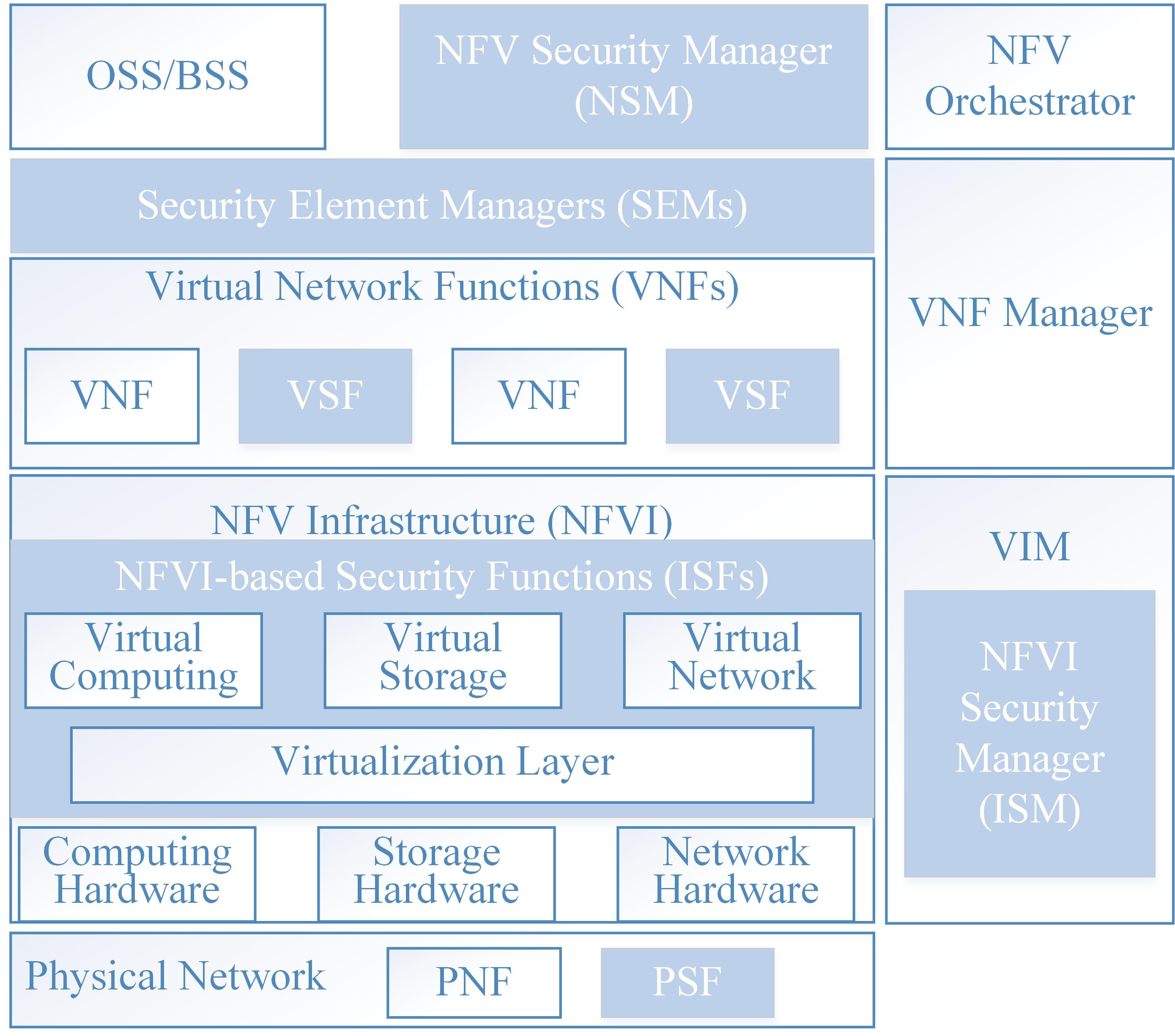}               
  \caption{High-level NFV security management framework.}
	\label{Fig4}
\end{figure}

ETSI proposes a high-level security management framework~\cite{ETSI2}, as shown in Fig.~\ref{Fig4}, to meet these requirements. From top, the NFV Security Manager (NSM) copes with complexity, separation of domains, and consistency challenges for security management of network services. Security Element Managers (SEMs) manage different security functions. Tailored security functions are implemented as VNFs called Virtual Security Functions (VSFs). VSF can be a firewall, Intrusion Detection/Preventions System (IDS/IPS), etc., and can be used to protect other VNFs as well. A security function provided by the NFV Infrastructure (NFVI) is called NFVI-based Security Function (ISF) that can include software, hardware or virtual security systems. Part of the non-virtualized traditional network, Physical Security Function (PSF), is the hybrid (virtual and non-virtual) network and is managed by SEM instead of VIM. PSF is added to provide full security; however, it is not part of the fully virtualized environment.

The overall security management is provided by NSM which is also involved in security policy planning. The NFVI Security Manager (ISM) is a security management function in the NFVI layer that builds and manages security in NFVI to support NSM request for managing security of network services in higher layer. However, there should be security controls and security policies, and clear security principles defining privileges for different functions. The security monitoring of VMNs will involve monitoring of management, services and systems of VMNs. The management security monitoring include monitoring of attacks, deployed security policies, and monitoring of operation behavior of the environment. The service security monitoring includes monitoring interfaces and handling of service procedures (e.g., signaling). The system security monitoring has many prongs such as monitoring system integrity, logs, traffic, resource usage, and security management processes. The monitoring techniques can be either passive, active or a combination of both. However, security monitoring of dynamic VNFs will be challenging. For example, tracking the traffic of moving functions and services will require synchronizing different network systems and functions. Furthermore, trust establishment will be very important in VMNs as described below.


\subsection{Trust Establishment}
Trust and privacy will be primary concerns in shared environments.Trust in communication networks is about the expected outcomes of communicating with remote entities. Trusted networking encompasses questions of losing data or assets, network resources, and privacy during communication~\cite{ylianttila20206g}. In VMNs, trust establishment can be rather tricky mainly because of sophisticated tools over the network used to hide identities. One of the basic approaches to ensure trust over the network is strong identity binding techniques starting from the locator/ID split of communicating devices~\cite{9083917}. Raimo Kantola in~\cite{9083917} described the potential pitfalls and possible mitigation principles and techniques with great detail for 5G/6G. In VMNs, the case is same with the only exception that tracking in virtualized environment will be more challenging than the physical counterpart. 

\subsection{Privacy}

Any information from which a person or attributes of a person can be identified must be kept private. Generally, virtual networks support privacy since information from a big number of people is not directly linked to anyone person. There are even methods that create virtual users to enhance privacy of individuals over communication networks as presented in~\cite{shiloh2008method}. However, in VMNs the control of user over his information is much lower, and can be barely traced to know where the information actually resides. This links privacy of users in VMNs to privacy in cloud systems. There are a number of approaches that can be used to secure privacy in virtual systems in clouds such as described in~\cite{kumar2018user, 7364212, 9004559}.

\subsection{Standardization Efforts}
The 3GPP working group, i.e., SA WG3~\cite{zhang2017overview} is monitoring security of 5G, including virtualization and NFV, etc. ETSI is more fucused on NFV and virtualization. Thus, the ETSI Industry Specification Group for NFV~\cite{ETSI1} is working on security with a dedicated group called the ISG NFV Security group (ISG NFV Sec). The latest, 2019-2020, NFV release 4 covers the verification, and certification procedures and mechanisms. The ISG NFV Sec group has published several group specification documents related to security such as access token specification for API access~\cite{ETSI3}, VNF Package Security Specification~\cite{ETSI4}, Security Specification for MANO Components and Reference points~\cite{ETSI5}, report on NFV Remote Attestation Architecture~\cite{ETSI6}, and on security management and monitoring specification in release 3~\cite{ETSI7}. Furthermore, there are several other reports on privacy, regulations, and trust guidance.

\section{Open Research Areas}
There are many open research areas in securing VMNs. Since the deployment is very limited, more security concerns will arise as we move forward towards its practical use. The potential challenges and open research areas can be grasped from Table III. The challenges in Table III, based in ITU-T security recommendations, are listed from low (L) to medium (M) and high (H). The challenges are more threatening where there are few resources with respect to the security implication. For example, access control is more threatening in SDN since unauthorized access to the SDN controller can lead to a hijack of the whole network. Yet, the control platforms are not cable to have heavy security mechanisms due to scalability challenges, as compared to centralized cloud platforms. Therefore, access control will have higher security implications in SDN than cloud platforms. The challenges are labeled low, where there is no direct implication of the particular challenge on the technology. For example, availability of a VNF can be a security challenge, yet VNFs can be created and moved around different resources at run-time without compromising on running flows. Therefore, it is considered as a lower (L) challenge, even though in the ultimate sense every challenge must be considered as high, e.g., for trust.

\begin{table} [ht]
\scriptsize
\renewcommand{\arraystretch}{1.0} 
\caption{Security domain implications and challenge level in each technology}
\label{table_layer3} 
\centering 
\begin{tabular}{|p{3 cm} |p{0.5 cm} |p{0.5 cm} |p{0.5 cm} | p{0.5 cm} |p{0.75 cm} |} 
\hline 
\bfseries \multirow{2}{*}{ Security Domain} & \multicolumn{4}{|c|} {\textbf{Technologies}} & \bfseries  \multirow{2}{*}{Trust} \ \\ [0.5ex] \cline{2-5}

 &SDN  &  cloud &   VMs &  VNF&\\  \hline

Access Control  & H & M & M & M & H\\ \hline
Authentication  & H & H & H & H & H \\ \hline
Non-Repudiation&   M & M & L & L & H\\ \hline
Data Confidentiality  & L& H &L & L & H \\ \hline
Communication security   & H & L&L &L  & H \\ \hline
Data integrity   &  L & H & L & L  & H \\ \hline
Availability  &  H & M & L & L & H \\ \hline
\textbf{Privacy} & L & H & L & L & H  \\ \hline

\hline

\end{tabular}
\vspace{-4mm}
\end{table}

\section{Conclusions}

VMNs will share the same physical infrastructure with many operators including MNOs and VMNOs. Therefore, the security environment will be complex, and dependent on security of enabling technologies and other operators. New security concepts must be brought forth that can secure both MNOs and VMNs at the same time. Strong isolation techniques, secure management, and fast and efficient monitoring systems will play an important role. Since, the deployment of VMNs at a large level is very limited, the security threats cannot be fully realized. However, security-by-design will be the key to secure and safe operation of future VMNs.  

\ifCLASSOPTIONcaptionsoff
  \newpage
\fi

\bibliographystyle{./bibliography/IEEEtran}
\bibliography{./bibliography/IEEEabrv,./bibliography/IEEEexample}

\begin{thebibliography}{10}
\providecommand{\url}[1]{#1}
\csname url@samestyle\endcsname
\providecommand{\newblock}{\relax}
\providecommand{\bibinfo}[2]{#2}
\providecommand{\BIBentrySTDinterwordspacing}{\spaceskip=0pt\relax}
\providecommand{\BIBentryALTinterwordstretchfactor}{4}
\providecommand{\BIBentryALTinterwordspacing}{\spaceskip=\fontdimen2\font plus
\BIBentryALTinterwordstretchfactor\fontdimen3\font minus
  \fontdimen4\font\relax}
\providecommand{\BIBforeignlanguage}[2]{{%
\expandafter\ifx\csname l@#1\endcsname\relax
\typeout{** WARNING: IEEEtran.bst: No hyphenation pattern has been}%
\typeout{** loaded for the language `#1'. Using the pattern for}%
\typeout{** the default language instead.}%
\else
\language=\csname l@#1\endcsname
\fi
#2}}
\providecommand{\BIBdecl}{\relax}
\BIBdecl

\bibitem{8703474}
Y.~{Xiao}, M.~{Krunz}, and T.~{Shu}, ``Multi-operator network sharing for
  massive iot,'' \emph{IEEE Communications Magazine}, vol.~57, no.~4, pp.
  96--101, 2019.

\bibitem{YI2018212}
B.~Yi, X.~Wang, K.~Li, S.~k.~Das, and M.~Huang, ``A comprehensive survey of
  network function virtualization,'' \emph{Computer Networks}, vol. 133, pp.
  212 -- 262, 2018.

\bibitem{CERRATO2015380}
I.~Cerrato, A.~Palesandro, F.~Risso, M.~Suñé, V.~Vercellone, and H.~Woesner,
  ``Toward dynamic virtualized network services in telecom operator networks,''
  \emph{Computer Networks}, vol.~92, pp. 380 -- 395, 2015, software Defined
  Networks and Virtualization.

\bibitem{7045396}
B.~{Han}, V.~{Gopalakrishnan}, L.~{Ji}, and S.~{Lee}, ``{Network function
  virtualization: Challenges and opportunities for innovations},'' \emph{IEEE
  Communications Magazine}, vol.~53, no.~2, pp. 90--97, Feb 2015.

\bibitem{7194059}
{Costa-Requena, et al.}, ``{{SDN and NFV integration in generalized mobile
  network architecture}},'' in \emph{{Networks and Communications (EuCNC), 2015
  European Conference on}}, {June} {2015}, pp. {154--158}.

\bibitem{8334918}
I.~{Ahmad}, T.~{Kumar}, M.~{Liyanage}, J.~{Okwuibe}, M.~{Ylianttila}, and
  A.~{Gurtov}, ``{Overview of 5G Security Challenges and Solutions},''
  \emph{IEEE Communications Standards Magazine}, vol.~2, no.~1, pp. 36--43,
  2018.

\bibitem{8712553}
I.~{Ahmad}, S.~{Shahabuddin}, T.~{Kumar}, J.~{Okwuibe}, A.~{Gurtov}, and
  M.~{Ylianttila}, ``{Security for 5G and Beyond},'' \emph{IEEE Communications
  Surveys Tutorials}, vol.~21, no.~4, pp. 3682--3722, Fourthquarter 2019.

\bibitem{7901481}
G.~{Biczok}, M.~{Dramitinos}, L.~{Toka}, P.~E. {Heegaard}, and
  H.~{Lonsethagen}, ``Manufactured by software: Sdn-enabled multi-operator
  composite services with the 5g exchange,'' \emph{IEEE Communications
  Magazine}, vol.~55, no.~4, pp. 80--86, 2017.

\bibitem{7226783}
{I. Ahmad and S. Namal and M. Ylianttila and A. Gurtov}, ``{{Security in
  Software Defined Networks: A Survey}},'' \emph{{IEEE Communications Surveys
  Tutorials}}, vol.~{17}, no.~{4}, pp. {2317--2346}, {Fourthquarter} {2015}.

\bibitem{azzouni2016fingerprinting}
A.~Azzouni, O.~Braham, T.~M.~T. Nguyen, G.~Pujolle, and R.~Boutaba,
  ``Fingerprinting openflow controllers: The first step to attack an sdn
  control plane,'' in \emph{2016 IEEE Global Communications Conference
  (GLOBECOM)}.\hskip 1em plus 0.5em minus 0.4em\relax IEEE, 2016, pp. 1--6.

\bibitem{7480416}
H.~{Cui}, G.~O. {Karame}, F.~{Klaedtke}, and R.~{Bifulco}, ``On the
  fingerprinting of software-defined networks,'' \emph{IEEE Transactions on
  Information Forensics and Security}, vol.~11, no.~10, pp. 2160--2173, 2016.

\bibitem{7399014}
M.~{Liyanage}, I.~{Ahmad}, M.~{Ylianttila}, A.~{Gurtov}, A.~B. {Abro}, and
  E.~M. {de Oca}, ``Leveraging lte security with sdn and nfv,'' in \emph{2015
  IEEE 10th International Conference on Industrial and Information Systems
  (ICIIS)}, 2015, pp. 220--225.

\bibitem{hayes2008cloud}
B.~Hayes, ``Cloud computing,'' 2008.

\bibitem{6963805}
Y.~{Cai}, F.~R. {Yu}, and S.~{Bu}, ``Cloud computing meets mobile wireless
  communications in next generation cellular networks,'' \emph{IEEE Network},
  vol.~28, no.~6, pp. 54--59, 2014.

\bibitem{7807196}
M.~{Satyanarayanan}, ``The emergence of edge computing,'' \emph{Computer},
  vol.~50, no.~1, pp. 30--39, 2017.

\bibitem{8016573}
Y.~{Mao}, C.~{You}, J.~{Zhang}, K.~{Huang}, and K.~B. {Letaief}, ``A survey on
  mobile edge computing: The communication perspective,'' \emph{IEEE
  Communications Surveys Tutorials}, vol.~19, no.~4, pp. 2322--2358, 2017.

\bibitem{7931566}
T.~{Taleb}, K.~{Samdanis}, B.~{Mada}, H.~{Flinck}, S.~{Dutta}, and
  D.~{Sabella}, ``On multi-access edge computing: A survey of the emerging 5g
  network edge cloud architecture and orchestration,'' \emph{IEEE
  Communications Surveys Tutorials}, vol.~19, no.~3, pp. 1657--1681,
  thirdquarter 2017.

\bibitem{7513863}
M.~{Peng}, S.~{Yan}, K.~{Zhang}, and C.~{Wang}, ``Fog-computing-based radio
  access networks: issues and challenges,'' \emph{IEEE Network}, vol.~30,
  no.~4, pp. 46--53, 2016.

\bibitem{8951281}
C.~{Li}, Y.~{Lin}, Y.~{Lai}, H.~{Chien}, Y.~{Huang}, P.~{Huang}, and H.~{Liu},
  ``Transparent aaa security design for low-latency mec-integrated cellular
  networks,'' \emph{IEEE Transactions on Vehicular Technology}, vol.~69, no.~3,
  pp. 3231--3243, 2020.

\bibitem{9004559}
P.~{Zhao}, H.~{Huang}, X.~{Zhao}, and D.~{Huang}, ``P3: Privacy-preserving
  scheme against poisoning attacks in mobile-edge computing,'' \emph{IEEE
  Transactions on Computational Social Systems}, vol.~7, no.~3, pp. 818--826,
  2020.

\bibitem{6897914}
A.~{Checko}, H.~L. {Christiansen}, Y.~{Yan}, L.~{Scolari}, G.~{Kardaras}, M.~S.
  {Berger}, and L.~{Dittmann}, ``Cloud ran for mobile networks—a technology
  overview,'' \emph{IEEE Communications Surveys Tutorials}, vol.~17, no.~1, pp.
  405--426, Firstquarter 2015.

\bibitem{7954591}
F.~{Tian}, P.~{Zhang}, and Z.~{Yan}, ``A survey on c-ran security,'' \emph{IEEE
  Access}, vol.~5, pp. 13\,372--13\,386, 2017.

\bibitem{ahmad20195}
I.~Ahmad, J.~Suomalainen, and J.~Huusko, ``5 g-core network security,''
  \emph{Wiley 5G Ref: The Essential 5G Reference Online}, pp. 1--18, 2019.

\bibitem{liyanage2018comprehensive}
M.~Liyanage, I.~Ahmad, A.~B. Abro, A.~Gurtov, and M.~Ylianttila, \emph{A
  Comprehensive Guide to 5G Security}.\hskip 1em plus 0.5em minus 0.4em\relax
  John Wiley \& Sons, 2018.

\bibitem{8806681}
E.~{Marku}, G.~{Biczók}, and C.~{Boyd}, ``Towards protected vnfs for
  multi-operator service delivery,'' in \emph{2019 IEEE Conference on Network
  Softwarization (NetSoft)}, 2019, pp. 19--23.

\bibitem{7502434}
W.~{Yang} and C.~{Fung}, ``A survey on security in network functions
  virtualization,'' in \emph{2016 IEEE NetSoft Conference and Workshops
  (NetSoft)}, June 2016, pp. 15--19.

\bibitem{NFVOrchSec}
M.~Pattaranantakul, Y.~Tseng, R.~He, Z.~Zhang, and A.~Meddahi, ``{A First Step
  Towards Security Extension for NFV Orchestrator},'' in \emph{Proceedings of
  the ACM International Workshop on Security in Software Defined Networks \&
  Network Function Virtualization}.\hskip 1em plus 0.5em minus 0.4em\relax New
  York, NY, USA: Association for Computing Machinery, 2017, p. 25–30.

\bibitem{tosca2015tosca}
O.~TOSCA, ``Tosca simple profile for network functions virtualization (nfv)
  version 1.0,'' 2015.

\bibitem{7014200}
J.~{Keeney}, S.~v.~d. {Meer}, and L.~{Fallon}, ``Towards real-time management
  of virtualized telecommunication networks,'' in \emph{10th International
  Conference on Network and Service Management (CNSM) and Workshop}, Nov 2014,
  pp. 388--393.

\bibitem{7113220}
R.~{Szabo}, M.~{Kind}, F.~{Westphal}, H.~{Woesner}, D.~{Jocha}, and
  A.~{Csaszar}, ``Elastic network functions: opportunities and challenges,''
  \emph{IEEE Network}, vol.~29, no.~3, pp. 15--21, May 2015.

\bibitem{ETSI2}
\BIBentryALTinterwordspacing
{{Network Functions Virtualisation (NFV) Release 3; Security; Security
  Management and Monitoring specification}}. [Online]. Available:
  \url{{https://www.etsi.org}}
\BIBentrySTDinterwordspacing

\bibitem{ylianttila20206g}
M.~Ylianttila, R.~Kantola, A.~Gurtov, L.~Mucchi, I.~Oppermann, Z.~Yan, T.~H.
  Nguyen, F.~Liu, T.~Hewa, M.~Liyanage \emph{et~al.}, ``6g white paper:
  Research challenges for trust, security and privacy,'' \emph{arXiv preprint
  arXiv:2004.11665}, 2020.

\bibitem{9083917}
R.~{Kantola}, ``{Trust Networking for Beyond 5G and 6G},'' in \emph{2020 2nd 6G
  Wireless Summit (6G SUMMIT)}, 2020, pp. 1--6.

\bibitem{shiloh2008method}
D.~Shiloh, ``Method and system for securing user identities and creating
  virtual users to enhance privacy on a communication network,'' Aug.~12 2008,
  uS Patent 7,412,422.

\bibitem{kumar2018user}
T.~Kumar, M.~Liyanage, I.~Ahmad, A.~Braeken, and M.~Ylianttila, ``User privacy,
  identity and trust in 5g,'' \emph{A Comprehensive Guide to 5G Security}, pp.
  267--279, 2018.

\bibitem{7364212}
L.~{Xu}, J.~{Lee}, S.~H. {Kim}, Q.~{Zheng}, S.~{Xu}, T.~{Suh}, W.~W. {Ro}, and
  W.~{Shi}, ``Architectural protection of application privacy against software
  and physical attacks in untrusted cloud environment,'' \emph{IEEE
  Transactions on Cloud Computing}, vol.~6, no.~2, pp. 478--491, 2018.

\bibitem{zhang2017overview}
X.~Zhang, A.~Kunz, and S.~Schr{\"o}der, ``{Overview of 5G security in 3GPP},''
  in \emph{2017 IEEE Conference on Standards for Communications and Networking
  (CSCN)}.\hskip 1em plus 0.5em minus 0.4em\relax IEEE, 2017, pp. 181--186.

\bibitem{ETSI1}
\BIBentryALTinterwordspacing
{{Network Functions Virtualisation. European Telecommunications Standards
  Institute (ETSI)}}. [Online]. Available:
  \url{{http://www.etsi.org/technologies-clusters/technologies/nfv}}
\BIBentrySTDinterwordspacing

\bibitem{ETSI3}
\BIBentryALTinterwordspacing
{{Network Functions Virtualisation (NFV) Release 2; Security; Access Token
  Specification for API Access}}. [Online]. Available:
  \url{{https://www.etsi.org}}
\BIBentrySTDinterwordspacing

\bibitem{ETSI4}
\BIBentryALTinterwordspacing
{{Network Functions Virtualisation (NFV) Release 2; Security; VNF Package
  Security Specification}}. [Online]. Available: \url{{https://www.etsi.org}}
\BIBentrySTDinterwordspacing

\bibitem{ETSI5}
\BIBentryALTinterwordspacing
{{Network Functions Virtualisation (NFV) Release 3; Security; Security
  Specification for MANO Components and Reference points}}. [Online].
  Available: \url{{https://www.etsi.org}}
\BIBentrySTDinterwordspacing

\bibitem{ETSI6}
\BIBentryALTinterwordspacing
{{ Network Functions Virtualisation (NFV); Security; Report on NFV Remote
  Attestation Architecture}}. [Online]. Available: \url{{https://www.etsi.org}}
\BIBentrySTDinterwordspacing

\bibitem{ETSI7}
\BIBentryALTinterwordspacing
{{Network Functions Virtualisation (NFV) Release 3; Security; System
  architecture specification for execution of sensitive NFV components}}.
  [Online]. Available: \url{{https://www.etsi.org}}
\BIBentrySTDinterwordspacing

\end{thebibliography}

\end{document}